\documentclass[preprint,12pt]{aastex}

\providecommand{\etal}{et~al.}
\providecommand{\mb}{$\Delta m_{15}(B)$}
\providecommand{\sw}{\sl Swift\/}
\providecommand{\bmax}{$B_{max}$}

\shorttitle{Swift Observations of SN2005am}
\shortauthors{Brown \etal}

\begin{document}


\title{Ultraviolet, Optical, and X-Ray Observations of the Type Ia Supernova 2005am with {\sl Swift\/}}

\author{P.~J.~Brown\altaffilmark{1,2},
        S.~T.~Holland\altaffilmark{3,4},
        C.~James\altaffilmark{5},
	P.~Milne\altaffilmark{6},
        P.~W.~A.~Roming\altaffilmark{1},
	K.~O.~Mason\altaffilmark{5}
	K.~L.~Page\altaffilmark{7},
        A.~P.~Beardmore\altaffilmark{7},
        D.~Burrows\altaffilmark{1},
        A.~Morgan\altaffilmark{1},
        C.~Gronwall\altaffilmark{1},
	A.~J.~Blustin\altaffilmark{5},
	P.~Boyd\altaffilmark{3},
	M.~Still\altaffilmark{3},
	A.~Breeveld\altaffilmark{5},
	M.~de~Pasquale\altaffilmark{5},
	S.~Hunsberger\altaffilmark{1},
	M.~Ivanushkina\altaffilmark{1},
	W.~Landsman\altaffilmark{3},
	K.~McGowan\altaffilmark{5},
	T.~Poole\altaffilmark{5},
	S.~Rosen\altaffilmark{5},
	P.~Schady\altaffilmark{5}, \&
        N.~Gehrels\altaffilmark{3}}

\altaffiltext{1}{Pennsylvania State University,
                 Department of Astronomy \& Astrophysics,
                 University Park, PA 16802}
                 
\altaffiltext{2}{pbrown@astro.psu.edu}                

\altaffiltext{3}{Laboratory for High Energy Astrophysics,
                 Goddard Space Flight Center,
                 Greenbelt, MD 20771}

\altaffiltext{4}{Universities Space Research Association}

\altaffiltext{5}{Mullard Space Science Laboratory,
		 Department of Space and Climate Physics,
		 University College London, Holmbury St Mary,
		 Dorking, Surrey, RH5 6NT,
		 UK}

\altaffiltext{6}{Steward Observatory,
		 933 N. Cherry Ave., RM N204
		 Tucson, AZ 85721
		 U.S.A.}

\altaffiltext{7}{Department of Physics \& Astronomy,
                 University of Leicester,
                 Leicester, LE1 7RH,
                 UK}


\begin{abstract}

     We present ultraviolet and optical light curves in six broadband
filters and grism spectra obtained by {\sl Swift\/}'s
Ultraviolet/Optical Telescope for the Type Ia supernova SN2005am.
The data were collected beginning about four days before the $B$-band
maximum, with excellent coverage of the rapid decline phase and later
observations extending out to 69 days after the peak.  The optical and near UV
light curve match well those of SN1992A.
The other UV observations constitute the first set of light curves shorter than 
2500 {\AA} and allow us to compare the light curve evolution in
three UV bands.  The UV behavior of this and other low redshift
supernovae can be used to constrain theories of progenitor evolution
or to interpret optical light curves of high redshift supernovae.
Using {\sl Swift\/}'s $X$-Ray Telescope, we also report the upper
limit to SN2005am's $X$-ray luminosity to be $1.77 \times 10^{40}$ erg
s$^{-1}$ in the 0.3--10 keV range from 58\,117 s of exposure time.

\end{abstract}


\keywords{supernovae: individual (SN2005am)}


\section{Introduction\label{SECTION:intro}}

     Type Ia supernovae (SNe) are among the brightest of astrophysical
events, making them useful as probes of the distant universe.  Because
Type Ia have similar luminosities at their peak, and a
well-established relationship between their peak brightness and rate
of decay, they are excellent standardizable candles.  The dispersion
in absolute magnitudes can be reduced to approximately 15\% by
calibrating the peak luminosity to other observable parameters such as
the $B$-band decline rate {\mb} \citep{P93,P99}.  SNe Ia gave the
first evidence that the expansion of the universe is accelerating
\citep{R98}, and they are used to constrain certain cosmological
parameters \citep{Perl99}

     UV observations are important for understanding the behaviour of
SNe.  For local ($z \approx 0$) SNe, observations in the UV can be used
to distinguish between different explosion models, as the UV emission 
probes the metallicity of the progenitor, as well as the degree of 
mixing of the synthesized $^{56}$Ni \citep{BS00}.  For
SNe observed at high redshifts, the rest-frame UV is redshifted into
the optical bands, so understanding the rest-frame UV behaviour of SNe
is critical to understanding the nature of more distant SNe Ia.

     Unfortunately, observations of local SNe in their rest-frame UV
are limited because they require space-based observatories.  The
International Ultraviolet Explorer (IUE) obtained UV spectroscopy of
12 Type Ia SNe, and the {\sl Hubble Space Telescope\/} ({\sl HST\/})
continues to obtain valuable UV spectroscopy and photometry (see \citealp{puv03} 
for a review of SN observations in the UV).  One of the
best observed Ia SN is 1992A which was observed by both IUE and HST
\citep{K93}.  There is also a Cycle 13 {\sl HST\/} programme (PI:
Filippenko) that has obtained UV observations of SN2004dt, SN2004ef, and
SN2005M (Wang, in prep.).  A larger sample is needed to determine how
uniform Type Ia SNe are in the UV.

     SN2005am was discovered by R.\ \citet{M05} in images from 2005 February 22 and 24 
(all dates UT).  The reported position was RA $=
9^\mathrm{h}16^\mathrm{m}12\fs47$ and Dec $=
-16\arcdeg18\arcmin16\arcsec$ (J2000) in NGC~2811 ($z = 0.007899$, \citealp{The98}).
It was confirmed a week later by K.\ Itagaki, by which time it had
brightened by about 3.5 magnitudes \citep{Y05}.  \citet{MKC05}
classified it from spectra as an early Type Ia SN on 2005 March
3, and the observations with the {\sl Swift\/} spacecraft reported
here began the next day.  In this paper we present UV and optical
photometry, grism spectroscopy, and an upper limit to the $X$-ray
luminosity.


\section{Observations and Reductions\label{SECTION:obs}}

     Observations of SN2005am were made with the {\sl Swift\/}
spacecraft \citep{G05} between 2005 Mar 4 and 2005 May 17.

     Swift's Ultraviolet/Optical Telescope (UVOT; \citealp{R05}) is a
30 cm telescope equipped with two grisms and six broadband filters.
The UV grism produces spectra from approximately
2000 {\AA} to 3400 {\AA}, and the V (optical) grism produces spectra 
from approximately 3000 {\AA} to 6000 {\AA}.  The filters and
their corresponding central wavelengths are UVW2 (1800 {\AA}), UVM2
(2200 {\AA}), UVW1 (2600 {\AA}), $U$ (3600 {\AA}), $B$ (4200 {\AA}),
and $V$ (5500 {\AA}).  To understand where in the spectrum of a SN
these filters correspond, Fig.~1 shows the filter transmission curves
superimposed on a combined spectrum of SN2005am using both the UV and
optical grisms.

     The $X$-Ray Telescope (XRT; \citealp{Bur05}) on {\sl Swift\/} is
sensitive to photons in the 0.2--10 keV energy range.  It has an
effective area of 110 cm$^2$ at 1.5 keV.  During the UVOT
observations, the XRT was gathering data in its Photon Counting mode.

     Observations began while the spacecraft was still in its
commissioning phase, and some calibrations are still ongoing.  Because
of this, the photometric zero points used will be explicitly given in
the event that future calibrations require an adjustment of those
values.


\subsection{Photometry\label{SECTION:phot}}

     SN2005am is located in the outer edge of the visible disc of
the host galaxy NGC~2811 and 
$6\farcs3$ from a ($V = 14.55 \pm 0.04$) star.  This star, and the
diffuse light from the host galaxy, complicate precision photometry.
High precision photometry will have to wait until the SN has
faded and the underlying diffuse light from NGC~2811 can be
subtracted.  It is anticipated that UVOT will reobserve NGC~2811 at
that time.

     We performed aperture photometry in a circular aperture with a
radius of $3\farcs5$ centered on the SN.  A sky annulus with an
inner radius of $20\arcsec$ and a width of $5\arcsec$ was used to
estimate the local background.  We found that these choices minimized 
contamination from the host galaxy and the nearby bright source while 
including most of the light from the supernova.

     The UVOT is a photon-counting detector, so it suffers from
coincidence losses for sources with high count rates.  This occurs
when the count rate approaches the frame rate and all photons are not
counted, and for very high count rates the number of photons missed
cannot be calculated.  This corresponds to saturation magnitudes of
$V_{\rm coinc} = 12.94$, $B_{\rm coinc} = 14.23$, $U_{\rm coinc} =
13.45$, $UVW1_{\rm coinc} = 12.93$, $UVM2_{\rm coinc} = 12.30$, and
$UVW2_{\rm coinc} = 12.93$.  Coincidence corrections were applied to
all of our data, though the nearby star may have caused the
coincidence losses to have been underestimated. At the SN's peak
brightness, our observations in the $B$ were saturated and should be
treated as lower limits to the true $B$-band magnitudes.

     The magnitudes were transformed to Vega magnitudes using Landolt
standards for the optical bands \citep{L92} and white dwarfs observed
by IUE for the UV bands \citep{WS91}.  The following preliminary
flight photometric zero points were applied: $ZP_V = 17.83 \pm 0.09$,
$ZP_B = 19.12 \pm 0.08$, $ZP_U = 18.34 \pm 0.16$, $ZP_{UVW1} = 17.82
\pm 0.23$, $ZP_{UVM2} = 17.19 \pm 0.26$, and $ZP_{UVW2} = 17.82 \pm
0.27$.  Color terms have not been calibrated, so they were not applied
to the photometry.  The photometry is presented in Table~1.

\subsection{Grism Spectroscopy \label{SECTION:spec}}

     Table~2 lists the grism exposures taken by UVOT\@.  The Epoch
column lists the number of days after maximum light.  The spectra of
SN2005am (integrated from both the UV and V grism exposures) were
extracted using the {\sc Swifttools} software included in the standard
HEASOFT software package.

     Care was taken to perform the background subtraction of the grism
data using clean regions, but this was hard to achieve due to the
crowding of the field.  Sequence IDs 00030010007 and 00030010011 were
particularly affected by this problem. All of the V grism observations
suffer from a high background level due to the neighbouring galaxy,
but, since both the background and the spectral regions are likewise
affected, the one should cancel the other. Contamination of the first
order spectra by close and overlapping zeroth order spectra adversely
affected the quality of the spectra from sequence IDs 00030010003 and
30010004. In order to circumvent the contamination in these regions a
narrow source width was specified for the spectral extraction from
sequence IDs 00030010003 and 00030010007. This may have resulted in some
loss of flux.

\subsection{X-Ray Data\label{SECTION:xray}}

     All XRT data between 2005 March 4 and April 22 (sequences
00030010001 to 00030010073) were consistently processed with {\sc
Swifttools} 2.0, using the default GTI-screening parameters. The
Photon Counting mode event-lists were then co-added, producing a total
exposure time of 58\,117 seconds.  No source was detected at the
coordinates of the SN, with a 3$\sigma$ upper limit of $5.4 \times
10^{-4}$ count s$^{-1}$, over the 0.2--10 keV band.


\section{Results\label{SECTION:results}}

     The light curves are presented in Fig.~2.  For comparison we have
overlaid template curves for the $B$ and $V$ bands from SN1992A
\citep{H96} and for the UVW1 band from SNe 1992A and 1990N
\citep{K93}.

\subsection{Optical Light Curves\label{SECTION:opt}}

     Because of the saturation in the $B$ band, the time of {\bmax}
and the value of {\mb} are not well determined by our data alone.  To
estimate their values, the SN2005am light curves in the $B$ and $V$
bands were compared to templates from \citet{H96}.  Due to the
scatter, our light curves are consistent with SN1992A ({\mb} = 1.47)
or the template of SNe 1992bo/1993H ({\mb} = 1.69).  Ground based data
confirm an intermediate value for {\mb} (M. Hamuy, private
communication).  We estimate that {\bmax} occured on JD $2453438 \pm
1$ day (2005 March 8).  Deviations from the templates at late times
most likely result from contamination by the host galaxy and the
nearby foreground star.  Subtracting template images of the host
galaxy after the SN has faded will improve the photometry most for
these later observations.

\subsection{Ultraviolet Light Curves\label{SECTION:uv}}

     The UV photometry presented here is a unique set.  The only two
other Type Ia SNe with UV photometry available in the literature are
SNe 1992A and 1990N.  Using those two SNe, \citet{K93} composed
a template light curve using {\sl HST\/}'s F275W filter and the flux
extracted from IUE spectra over a similar range, but cut off at 3400
{\AA}.  This band, named *F275W in the above paper, covers a similar
wavelength range as UVOT's UVW1 filter, so this template is compared
to the UVOT data in Fig.~3.  Our observations agree very well with the
template.  This is reassuring, since the optical light curves of
SNe 2005am and 1992A are similar.  SN1990N had a shallower decay in
the optical ({\mb} = 1.07;
\citealp{P99}), and in Fig.~3 of \citet{K93} appears to be slightly
broader, but the observations do not extend past four days after
{\bmax}.  UV observations using {\sl HST\/} or {\sw} of SNe with a
variety of values of {\mb} would be valuable in calibrating a peak
luminosity-width relationship in the UV which will be useful for
cosmology.

     Blueward of 2500 {\AA}, only a few data points are available for
SN1992A, which marginally matched a $U$-aband template \citep{K93}.
The UVW2 and UVM2 curves presented here are the first opportunity to
see the photometric evolution in this spectral range.  One feature is
that the decay rate in the UV is actually shallower than in the $U$
and $B$ bands.  To quantify this, a subsection of the photometry
beginning three days after maximum and extending to day twenty are
used to calculate the decay rate in each band.  The decay rates (in
magnitudes per day) over this period are $D_V = 0.0684 \pm 0.0008$,
$D_B = 0.135 \pm 0.001$, $D_U = 0.158 \pm 0.001$, $D_{UVW1} = 0.129
\pm 0.002$, $D_{UVM2} = 0.069 \pm 0.007$, and $D_{UVW2} = 0.114 \pm
0.003$.  An unexplained feature is that the decay rate in the UVM2 filter
is shallower than the decay rate in the UV filters overlapping with it
on both sides.

     UV observations of SNe such as these could be useful for
interpreting high redshift SNe observed in the optical bands.  For
example, \citet{K93} noted that the light curve of a nearby SNe
observed in the *F275W filter would be similar to that of a SN at $z =
0.6$ observed in the $B$ band.  The same would be true of UVOT's UVW1
filter.  Similarly, a rest frame SN observed with UVOT's UVW2 filter
should match a $B$ band observation of a SN at a redshift of $z
\approx 1.5$.  Time dilation would need to be taken into account, as
well as $K$ corrections which have not been done in the UV\@.

\subsection{Spectra\label{SECTION:specres}}

     The grism spectra are presented in Fig.~3.  The optical spectra
are typical of Type Ia SNe, featuring P-Cygni line profiles
superimposed on a roughly thermal continuum.  The UV continuum is
suppressed by deep absorption from blends of iron peak elements.
Broad peaks on either side of 3000 {\AA} resemble those seen in other
SNe observed in the UV \citep{Ben82} and identified by \citet{B86} as
blended lines of \ion{Fe}{2} and \ion{Co}{2}.  A more detailed
analysis of these spectra will be performed after better calibrations
are available.

\subsection{X-Ray Luminosity\label{section:xrl}}

At early epochs, hard $X$-rays are created in SNe Ia by gamma-rays, produced in 
the decays of $^{56}$Ni synthesized in the SN explosion. These gamma-rays 
scatter off of electrons in the ejecta via Compton scattering.  
The down-scattered photons can either 
escape the ejecta, be further down-scattered, or be absorbed via 
bound-free absorption. Further, energetic electrons created by the 
interaction of these gamma-ray photons slow in the ejecta, generating  
bremsstrahlung emission which produces photons in the 0.1 -- 100 keV 
energy range. $X$-rays in SNe are also caused by shocks of the expanding 
SN shell interacting with the nearby medium.  Type Ia SNe are not expected to
be bright in shock-induced $X$-ray emission because older, degenerate systems 
are expected to contain very little circumstellar material.  Assuming an 
$X$-ray spectrum similar to the Tycho SN remnant \citep{Dec01}, the upper limit
on SN2005am's count rate corresponds to an unabsorbed flux of $1.28
\times 10^{-13}$ erg s$^{-1}$ cm$^{-2}$ in the 0.3--10 keV range.  At
a redshift of 0.007899, this corresponds to an upper limit to the
unabsorbed luminosity in this range of $1.77 \times 10^{40}$ erg
s$^{-1}$.  In the 0.5--2 keV range the luminosity is less than $1.52
\times 10^{40}$ erg s$^{-1}$.  This is one to two orders of magnitude higher
than the luminosities predicted by Chandrasekhar and sub-Chandrasekhar 
models studied by Pinto, Eastman \& Rogers (2001). (That work included only 
$X$-rays created by the Compton scattering and bremsstrahlung mechanisms.)
The luminosity is also one order of magnitude higher than the upper limits 
for other young SNRs  studied by \citet{Bre03}, 
albeit at a much earlier observation time. 


\section{Conclusions\label{SECTION:conc}}

     SN2005am appears to be a normal Type Ia SN\@.  Its light curves
resemble those of SN1992A in both the optical and near UV\@.  The UV
and optical spectra resemble those of  other normal Type Ia SNe. 
No $X$-ray flux is detected from this SN. However, the expected flux 
levels from $^{56}$Ni-decay related keV emission, and from 
shock-induced emission are below the $X$-ray flux upper limits. 

     The observations reported here show the power of {\sl Swift}'s
     UVOT for studying the UV behavior of SNe.  UVOT can obtain
low-resolution grism spectra of bright SNe and follow their light
curve decay in three UV passbands.  Another of {\sw}'s unique
characteristics is its short term scheduling, allowing observations to
be made within days or even hours.  This allows UV observations to be
made at earlier epochs than with other UV instruments past or present.

     This paper presents a unique set of UV light curves for SN2005am
which are better sampled in time and cover a wider wavelength range
than any previous UV observations of a SN Ia. 
This SN was also well-observed in the optical and near-infrared 
wavelength ranges. Comparisons between the combined UV-OPT-NIR light curves 
of SN 2005am and radiation transport simulations of various theoretical 
explosion models for type Ia SNe will enhance our understanding of 
thermonuclear SN events. 

 Further {\sl Swift\/} UV observations of local Type Ia SNe
will help us understand the range in the UV properties of SNe Ia and
enhance their use as standardizable candles.  This in turn will allow
observations of high-redshift SNe Ia to better constrain the
cosmological parameters governing the expansion of the universe.


\acknowledgements

This work made use of the NASA/IPAC Extragalactic Database.



\begin{figure}
\epsscale{1}
\plotone{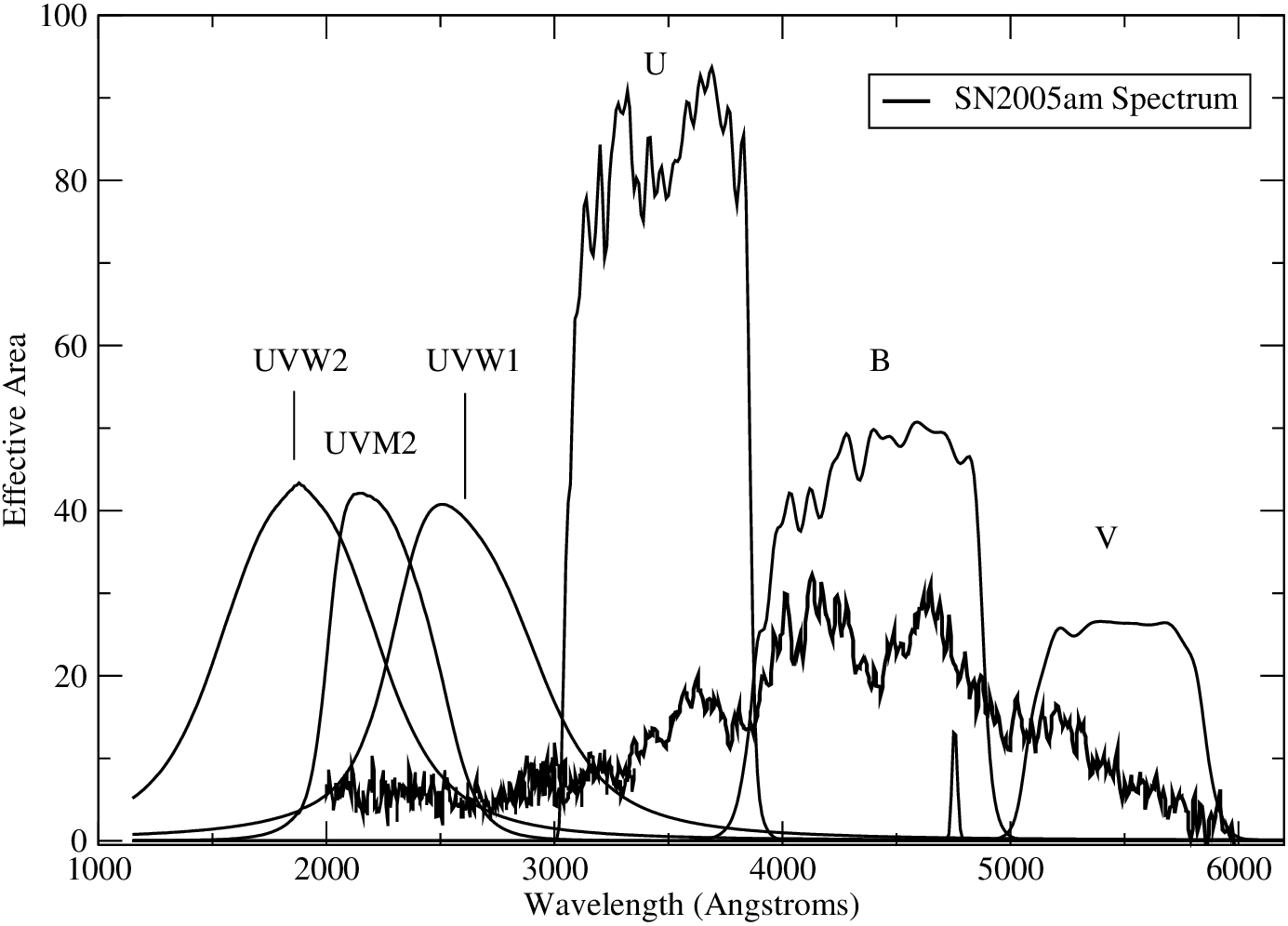}
\caption{Effective area curves of UVOT's six broadband filters 
superimposed on a spectrum of SN2005am.}.\label{fig1}
\end{figure}

\begin{figure}
\epsscale{1}
\plotone{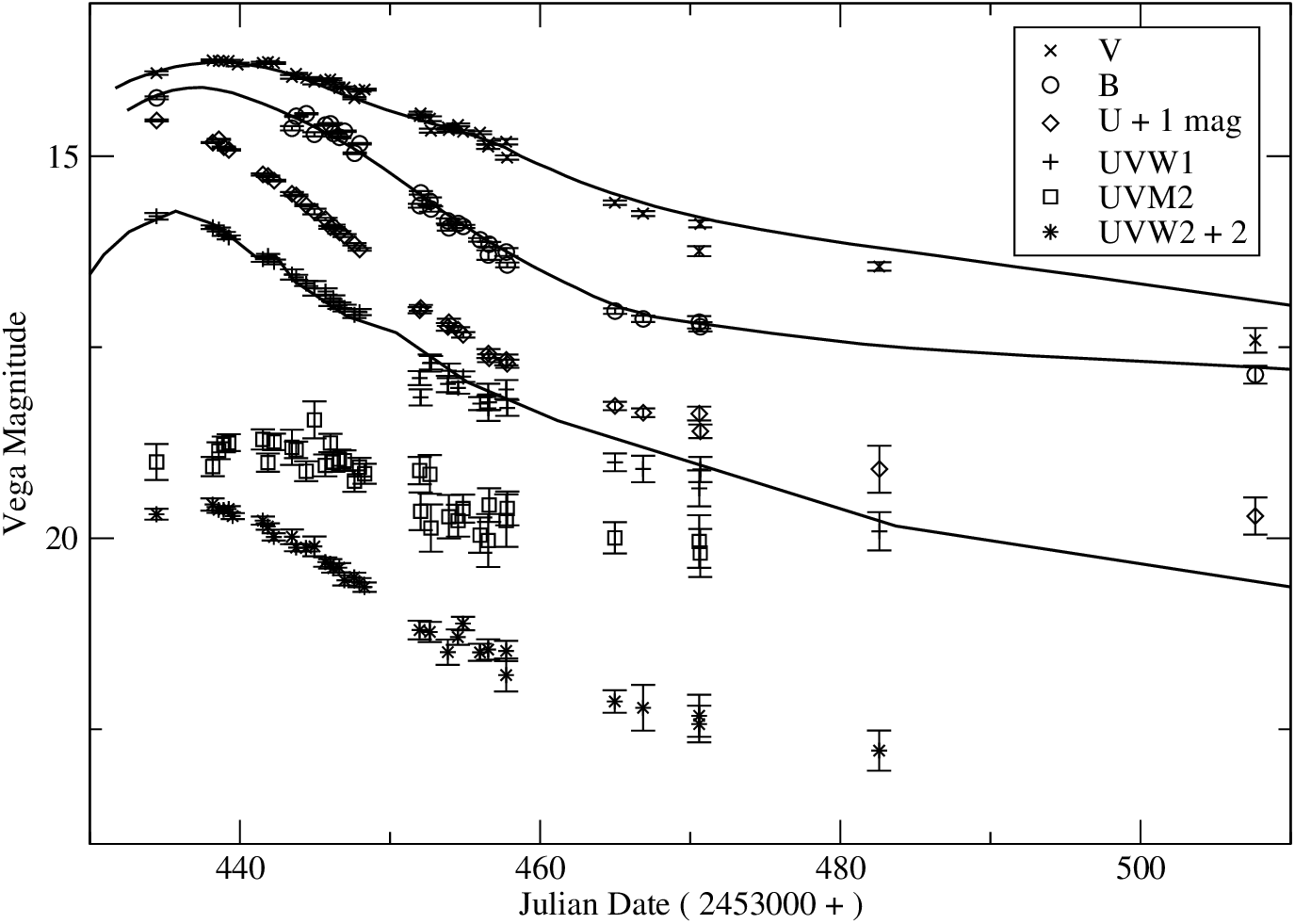}
\caption{Ultraviolet and Optical Light Curves Of SN2005am obtained by
UVOT\@.  For visual clarity, the U curve has been shift by + 1 magnitude and the UVW2 curve by + 2 magnitudes.  Overlaid are B and V templates of SN1992A and the
*F275W template from SNe 1992A/1990N (for our UVW1 curve)}.\label{fig2}
\end{figure}

\begin{figure}
\epsscale{1}
\plotone{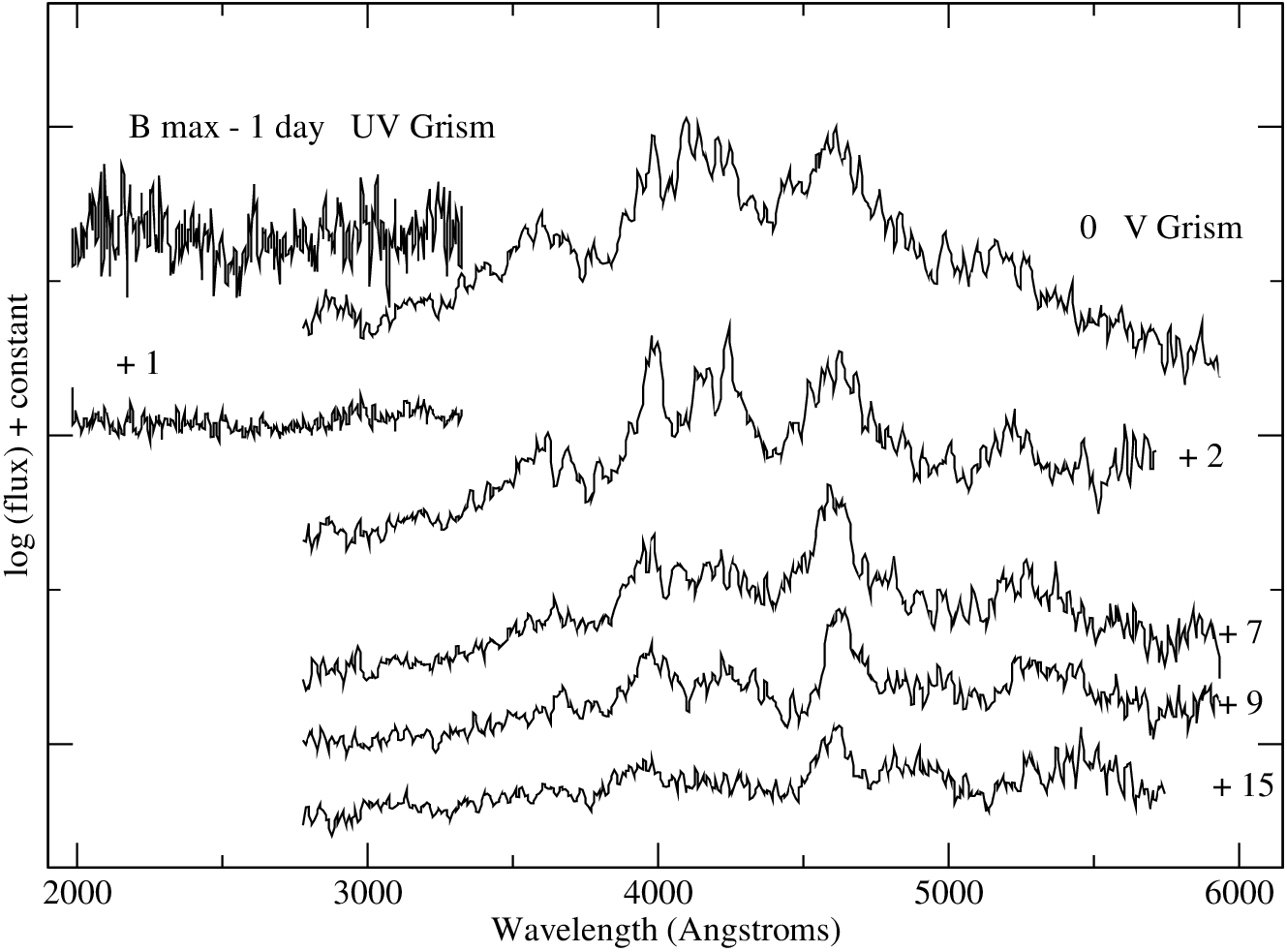}
\caption{UVOT Grism Spectra of SN2005am.  Each spectrum is labelled 
with its epoch in days from \bmax.The vertical scale is given in logarithmic 
units of ergs $s^{-1}$ $cm^{-2}$ $\AA^{-1}$.  The spectra have been trimmed from their 
full size to avoid contamination by other spectral orders.}.\label{fig3}
\end{figure}


\begin{deluxetable}{ccccccccccccc}

\rotate

\tabletypesize{\footnotesize}


\tablecaption{Optical and Ultraviolet Light Curves of SN2005am}

\tablenum{1}

\tablehead{\colhead{Julian Date} & \colhead{UVW2} & \colhead{$E_{UVW2}$} & \colhead{UVM2} & \colhead{$E_{UVM2}$} & \colhead{UVW1} & \colhead{$E_{UVW1}$} & \colhead{U} & \colhead{$E_{U}$} & \colhead{B} & \colhead{$E_{B}$} & \colhead{V} & \colhead{$E_{V}$} \\ 
\colhead{} & \colhead{} & \colhead{} & \colhead{} & \colhead{} & \colhead{} & \colhead{} & \colhead{} & \colhead{} & \colhead{} & \colhead{} & \colhead{} & \colhead{} } 

\startdata
2453434.442 & 17.69 & 0.072 & 19.00 & 0.237 & 15.79 & 0.043 & 13.53 & 0.013 & 14.24 & 0.019 & 13.91 & 0.023 \\
2453438.186 & 17.56 & 0.081 & 19.06 & 0.127 & 15.93 & 0.024 & 13.82 & 0.008 & $<14.23$ & 0.006 & 13.75 & 0.011 \\
2453438.586 & 17.63 & 0.038 & 18.86 & 0.110 & 15.96 & 0.024 & 13.78 & 0.008 & $<14.23$ & 0.006 & 13.76 & 0.010 \\
2453438.922 & 17.63 & 0.037 & 18.77 & 0.103 & 16.02 & 0.025 & 13.88 & 0.008 & $<14.23$ & 0.006 & 13.76 & 0.010 \\
2453439.257 & 17.62 & 0.040 & 18.75 & 0.110 & 16.06 & 0.026 & 13.919 & 0.009 & $<14.23$ & 0.005 & 13.75 & 0.011 \\
2453439.512 & 17.71 & 0.040 & \nodata & \nodata & \nodata & \nodata & \nodata & \nodata & \nodata & \nodata & \nodata & \nodata \\
2453439.851 & \nodata & \nodata & \nodata & \nodata & \nodata & \nodata & \nodata & \nodata & \nodata & \nodata & 13.80 & 0.024 \\
2453441.527 & 17.77 & 0.053 & 18.71 & 0.132 & 16.35 & 0.037 & 14.24 & 0.013 & $<14.23$ & 0.009 & 13.78 & 0.014 \\
2453441.868 & 17.85 & 0.042 & 19.01 & 0.119 & 16.31 & 0.028 & 14.26 & 0.010 & $<14.23$ & 0.006 & 13.77 & 0.010 \\
2453442.271 & 17.98 & 0.048 & 18.74 & 0.109 & 16.39 & 0.031 & 14.32 & 0.011 & $<14.23$ & 0.006 & 13.78 & 0.011 \\
2453443.463 & 17.98 & 0.096 & 18.81 & 0.230 & 16.55 & 0.067 & 14.49 & 0.025 & 14.64 & 0.028 & 13.96 & 0.025 \\
2453443.744 & 18.12 & 0.048 & 18.84 & 0.108 & 16.59 & 0.032 & 14.52 & 0.012 & 14.48 & 0.008 & 13.92 & 0.011 \\
2453444.414 & 18.13 & 0.048 & 19.12 & 0.129 & 16.67 & 0.034 & 14.64 & 0.013 & 14.45 & 0.008 & 13.98 & 0.012 \\
2453444.959 & 18.11 & 0.130 & 18.45 & 0.242 & 16.73 & 0.095 & 14.73 & 0.037 & 14.72 & 0.028 & 14.03 & 0.034 \\
2453445.69 & 18.31 & 0.061 & 19.05 & 0.143 & 16.77 & 0.041 & 14.82 & 0.017 & 14.59 & 0.010 & 14.01 & 0.014 \\
2453446.024 & 18.34 & 0.062 & 18.75 & 0.120 & 16.92 & 0.044 & 14.93 & 0.018 & 14.58 & 0.010 & 13.99 & 0.014 \\
2453446.224 & 18.39 & 0.058 & 19.01 & 0.125 & 16.86 & 0.039 & 14.91 & 0.016 & 14.70 & 0.010 & 14.05 & 0.013 \\
2453446.627 & 18.39 & 0.059 & 18.96 & 0.125 & 16.96 & 0.041 & 15.00 & 0.017 & 14.75 & 0.010 & 14.11 & 0.014 \\
2453446.962 & 18.55 & 0.067 & 18.98 & 0.133 & 16.99 & 0.044 & 15.04 & 0.018 & 14.68 & 0.010 & 14.10 & 0.014 \\
2453447.628 & 18.51 & 0.060 & 19.25 & 0.141 & 17.08 & 0.042 & 15.15 & 0.017 & 14.96 & 0.011 & 14.24 & 0.014 \\
2453447.965 & 18.58 & 0.060 & 19.07 & 0.122 & 17.04 & 0.040 & 15.22 & 0.018 & 14.84 & 0.009 & 14.14 & 0.013 \\
2453448.302 & 18.64 & 0.062 & 19.16 & 0.131 & \nodata & \nodata & \nodata & \nodata & \nodata & \nodata & 14.13 & 0.013 \\
2453448.302 & 18.64 & 0.062 & 19.16 & 0.131 & \nodata & \nodata & \nodata & \nodata & \nodata & \nodata & 14.13 & 0.013 \\
2453451.976 & 19.20 & 0.123 & 19.11 & 0.179 & 17.91 & 0.091 & 16.02 & 0.039 & 15.65 & 0.026 & 14.47 & 0.022 \\
2453452.044 & \nodata & \nodata & 19.65 & 0.247 & 18.16 & 0.105 & 15.99 & 0.039 & 15.48 & 0.023 & 14.44 & 0.022 \\
2453452.655 & 19.23 & 0.132 & 19.16 & 0.253 & 17.71 & 0.111 & \nodata & \nodata & 15.59 & 0.051 & 14.51 & 0.031 \\
2453452.725 & \nodata & \nodata & 19.87 & 0.310 & 17.70 & 0.088 & \nodata & \nodata & 15.69 & 0.034 & 14.67 & 0.027 \\
2453453.85 & 19.49 & 0.166 & $>20.13$ & 0.435 & 17.98 & 0.107 & 16.24 & 0.05 & 15.85 & 0.034 & 14.63 & 0.028 \\
2453453.918 & \nodata & \nodata & 19.72 & 0.284 & 17.81 & 0.095 & 16.17 & 0.047 & 15.94 & 0.035 & 14.66 & 0.027 \\
2453454.533 & 19.29 & 0.101 & 19.77 & 0.209 & 18.03 & 0.075 & 16.27 & 0.035 & 15.88 & 0.021 & 14.59 & 0.019 \\
2453454.867 & 19.12 & 0.088 & 19.61 & 0.183 & 17.89 & 0.069 & 16.34 & 0.035 & 15.92 & 0.020 & 14.68 & 0.019 \\
2453455.999 & 19.49 & 0.112 & 19.96 & 0.232 & 18.24 & 0.085 & \nodata & \nodata & 16.10 & 0.024 & 14.69 & 0.020 \\
2453456.541 & 19.46 & 0.133 & 20.03 & 0.347 & 18.15 & 0.173 & 16.58 & 0.058 & 16.29 & 0.062 & 14.88 & 0.031 \\
2453456.588 & \nodata & \nodata & \nodata & \nodata & 18.30 & 0.163 & 16.65 & 0.051 & 16.15 & 0.031 & 14.82 & 0.025 \\
2453456.6 & \nodata & \nodata & 19.56 & 0.220 & 18.22 & 0.100 & \nodata & \nodata & \nodata & \nodata & \nodata & \nodata \\
2453457.741 & 19.48 & 0.132 & 19.77 & 0.343 & 18.05 & 0.124 & 16.67 & 0.071 & 16.25 & 0.052 & 14.81 & 0.034 \\
2453457.806 & \nodata & \nodata & 19.61 & 0.221 & 18.29 & 0.104 & 16.72 & 0.053 & 16.42 & 0.031 & 15.02 & 0.028 \\
2453464.979 & 20.14 & 0.147 & 20.00 & 0.206 & 19.01 & 0.116 & 17.27 & 0.055 & 17.03 & 0.036 & 15.61 & 0.029 \\
2453466.858 & 20.22 & 0.299 & $>20.17$ & 0.566 & 19.09 & 0.173 & 17.36 & 0.057 & 17.13 & 0.041 & 15.75 & 0.032 \\
2453470.597 & 20.32 & 0.275 & 20.04 & 0.344 & 19.35 & 0.236 & 17.37 & 0.090 & 17.17 & 0.081 & 16.24 & 0.065 \\
2453470.656 & 20.43 & 0.240 & 20.19 & 0.315 & 19.10 & 0.165 & 17.60 & 0.088 & 17.23 & 0.060 & 15.89 & 0.046 \\
2453482.593 & 20.78 & 0.263 & $>20.71$ & 0.562 & 19.91 & 0.251 & 18.10 & 0.307 & \nodata & \nodata & 16.44 & 0.056 \\
2453507.639 & $>20.75$ & 0.868 & $>20.37$ & 1.822 & $>20.08$ & 4.513 & 18.71 & 0.244 & 17.86 & 0.117 & 17.41 & 0.161 \\
\enddata


\tablecomments{Errors quoted do not include any systematic error 
in the zeropoint or errors due to coincidence losses.  Lower limits 
to the brightness due to coincidence saturation and 3 $\sigma$ upper 
limits are indicated.}

\end{deluxetable}



\begin{deluxetable}{ccccc}

\tablecaption{UVOT Grism Observations}
\tablenum{2}
\tablehead{\colhead{Sequence ID} & \colhead{Grism Used} & 
\colhead{Date Time (UT)} & \colhead{Epoch} & \colhead{Exposure Length} \\ 
\colhead{} & \colhead{} & \colhead{} & \colhead{} & \colhead{} } 
\startdata
00030010003  & V & 2005-03-08 17:47:10 & -1 & 1812.7  \\
00030010004  & UV & 2005-03-07 23:07:34 & 0 & 2781.7  \\
00030010007  & UV & 2005-03-09 14:29:01 & +1 & 2369.9  \\
000300100011 & V & 2005-03-10 14:36:01 & +2 & 2332.4  \\
000300100024 & V & 2005-03-15 23:21:01 & +7 & 1828.9  \\
000300100035 & V & 2005-03-17 20:21:41 & +9 & 1828.3  \\
000300100054 & V & 2005-03-23 01:50:01 & +15 & 1838.5  \\
\enddata
\end{deluxetable}

\end{document}